\documentclass[twocolumn,showpacs,preprintnumbers]{revtex4}
\usepackage{amssymb}
\usepackage{amsfonts}
\usepackage{amsmath}
\usepackage{bm}

\input{tcilatex}

\begin{document}

\title{Small Entangled Quantum Worlds with a Simple Structure}
\author{E. D. Vol}
\email{vol@ilt.kharkov.ua}
\affiliation{B. Verkin Institute for Low Temperature Physics and Engineering of the
National Academy of Sciences of Ukraine, 47, Nauky Ave., Kharkov 61103,
Ukraine.}

\begin{abstract}
We introduce the notion of a small quantum world (SQW) , which in
our opinion, is very helpful in the situations when an
experimenter's tools for preparing quantum states and (or) for
measuring the observables of a quantum system under study are
restricted due to some kind of reasons. In this case it is advisable
to use as original an appropriate subspace of complete Hilbert space
of states and respectively to utilize observables acting only in
this subspace. If this subspace possesses some additional
symmetries, the structure of a pure states set, as a rule, is
simpler by far than in general case. Moreover in such SQWs some
specific irreducible entangled states may appear. The similar states
could be very helpful in various tasks connected with quantum
informational applications. In the present paper main ideas outlined
above are developed in detail in the simple and instructive case of
a two-qubit system in which the accessible space of states possesses
the additional symmetry structure of permutation group of three
elements.
\end{abstract}

\maketitle

It is well known that there are two fundamental concepts in quantum theory,
namely states and observables, and respectively an experimenter has to deal
with two main procedures that realize these concepts, namely preparation of
the required state and measuring the observable of interest in the end of
the experiment. Note that among all states of quantum system its pure
states, which contain the maximal possible information about the system,
play the most important role. One of the main postulates of quantum theory
claims that there is exact mapping between pure states of a quantum system
under study and vectors of appropriate Hilbert space. For example, in the
case of the most simple quantum system, that is a qubit, the relevant
Hilbert space, representing its states, is two-dimensional and there is a
geometrically descriptive image of the states of such system by means of the
Bloch sphere. As is well known, an arbitrary state of the qubit (mixed or
pure) with a density matrix ${\hat{\rho}}$ can be represented in the form ${%
\hat{\rho}}=\frac{1+\vec{P}\vec{\sigma}}{2}$, where $\hat{\sigma}_{i}$ $%
(i=1,2,3)$ are the Pauli matrices and $\vec{P}$ is appropriate Bloch vector.
The pure states of a qubit for which the condition ${\hat{\rho}}^{2}={\hat{%
\rho}}$ is satisfied are placed on the surface of the Bloch sphere with $%
\vec{P}=1$ while all the rest (mixed) states are settled inside this sphere.
Unfortunately, in more complex situations, in particular for composite
quantum systems the problem of description of the set of pure states by
geometrically distinct way is unsolved till now. Furthermore, when one is
operating with the states of composite systems, the important problem of
determination their entanglement ( the quantity that just specifies the
nonlocal informational resource and distinguishes quantum communicational
systems from classical ones), can be solved exactly also only for two-qubit
systems. In spite of enormous number of papers devoted to this problem (see
for example the comprehensive review of the topic \cite{1}), this problem
remains open up to now. In this connection we propose as a first step to
consider more simple problem, namely, to study some special classes of
quantum systems in which both the set of accessible quantum states and the
set of observables that are available for measurement are restricted by some
additional conditions.Evidently we will talking about specific subspaces of
the total Hilbert space and appropriate algebras of observables acting in
these subspaces. From physical point of view it means that, due to various
reasons, capabilities of an experimenter for the quantum states preparation
and for performing arbitrary measurements are restricted. Nevertheless,
since in this approach the space of accessible states of the system remains
linear and closed, all postulates of quantum theory continue to be valid.
Henceforth we will define as a small quantum world (SQW) certain subspace of
states within the complete Hilbert space with the relevant algebra of
observables acting in this subspace. The main goal of the present paper is
to demonstrate on particular examples that the structure of states in such
SQWs could be simpler by far than the structure of states in the enveloping
large quantum world. At first let us briefly remind one of the known
examples of SQW, namely, so called X-states in two-qubit quantum systems
\cite{2}. In this case one assumes that density matrix of any accessible
state of the quantum system under study can be represented as ${\hat{\rho}}=%
\begin{pmatrix}
\rho _{11} & 0 & 0 & \rho _{14} \\
0 & \rho _{22} & \rho _{23} & 0 \\
0 & \rho _{32} & \rho _{33} & 0 \\
\rho _{41} & 0 & 0 & \rho _{44}%
\end{pmatrix}%
$. The relevant basis of the algebra of observables acting on these states
consists of 8 operators (generators of the algebra). Let us write them
explicitly:1) $\hat{1}-$the unit operator ,2) another diagonal operator $%
\hat{E}=%
\begin{pmatrix}
1 & 0 & 0 & 0 \\
0 & -1 & 0 & 0 \\
0 & 0 & -1 & 0 \\
0 & 0 & 0 & 1%
\end{pmatrix}%
$, and in addition six operators $\hat{\lambda}_{i}$ and ${\hat{\tau}}_{i}$
\ ($i=1,2,3$) that are defined by analogy with the known Pauli
matrices,namely :
\begin{eqnarray*}
&&\hat{\lambda}_{1}=%
\begin{pmatrix}
0 & 0 & 0 & 1 \\
0 & 0 & 0 & 0 \\
0 & 0 & 0 & 0 \\
1 & 0 & 0 & 0%
\end{pmatrix}%
,\hat{\lambda}_{2}=%
\begin{pmatrix}
0 & 0 & 0 & -i \\
0 & 0 & 0 & 0 \\
0 & 0 & 0 & 0 \\
i & 0 & 0 & 0%
\end{pmatrix}%
, \\
&&\hat{\lambda}_{3}=%
\begin{pmatrix}
1 & 0 & 0 & 0 \\
0 & 0 & 0 & 0 \\
0 & 0 & 0 & 0 \\
0 & 0 & 0 & -1%
\end{pmatrix}%
\end{eqnarray*}%
and
\begin{eqnarray*}
&&{\hat{\tau}}_{1}=%
\begin{pmatrix}
0 & 0 & 0 & 0 \\
0 & 0 & 1 & 0 \\
0 & 1 & 0 & 0 \\
0 & 0 & 0 & 0%
\end{pmatrix}%
,{\hat{\tau}}_{2}=%
\begin{pmatrix}
0 & 0 & 0 & 0 \\
0 & 0 & -i & 0 \\
0 & i & 0 & 0 \\
0 & 0 & 0 & 0%
\end{pmatrix}%
, \\
&&{\hat{\tau}}_{3}=%
\begin{pmatrix}
0 & 0 & 0 & 0 \\
0 & 1 & 0 & 0 \\
0 & 0 & -1 & 0 \\
0 & 0 & 0 & 0%
\end{pmatrix}%
.
\end{eqnarray*}%
It is easy to point out the complete system of algebraic relations
connecting the generators of this algebra, namely:

\begin{eqnarray}
&&{\hat{\lambda}_{i}}{\hat{\lambda}_{j}}=\frac{1+\hat{E}}{2}\delta
_{ij}+i\varepsilon _{ijk}{\hat{\lambda}_{k}},  \notag  \label{1} \\
&&{\hat{\tau}_{i}}{\hat{\tau}_{j}}=\frac{1-\hat{E}}{2}\delta
_{ij}+i\varepsilon _{ijk}{\hat{\tau}_{k}},  \notag \\
&&{\hat{\lambda}_{i}}{\hat{\tau}_{j}}={\hat{\tau}_{j}}{\hat{\lambda}_{i}}=0,
\notag \\
&&\hat{E}{\hat{\lambda}_{i}}={\hat{\lambda}_{i}}\hat{E}={\hat{\lambda}_{i}}%
\text{ and }  \notag \\
&&\hat{E}{\hat{\tau}_{i}}={\hat{\tau}_{i}}\hat{E}=-{\hat{\tau}_{i}}
\end{eqnarray}

( where $\varepsilon _{ijk}$ is completely antisymmetric tensor).Any X-state
(that is its density matrix) can be represented as $\hat{\rho}=\frac{1+e\hat{%
E}+P_{i}{\hat{\lambda}_{i}}+S_{i}{\hat{\tau}_{i}}}{4}.$(where $e,P_{i},S_{i}$
are appropriate numerical coefficients). It is easy to see that 4
eigenvalues of such matrix can be calculated explicitly and are equal to: $%
\rho _{1,2}=\frac{1+e\pm \left\vert P\right\vert }{4}$ and $\rho _{3,4}=%
\frac{1-e\pm \left\vert S\right\vert }{4}.$ Evidently, that one can specify
two disjoint classes of pure X-states in such SQW: 1) with $e=1$ , $%
\left\vert P\right\vert =2,$ $\left\vert S\right\vert =0$ and 2) with $e=-1$
$\left\vert S\right\vert =2,$ $\left\vert P\right\vert =0$. Thus the
structure of pure states in this SQW turns out to be simpler by far than for
two-qubit states in general case. We will continue to study the properties
of X- states elsewhere, while the present paper will be devoted to another
interesting case, namely the SQWs that can be constructed on the basis of
the permutation group of four elements- $S_{4}$. It is well-known that due
to the principle of indistinguishability the permutation group plays the
fundamental and diverse role in quantum theory, but in this paper it will be
used only for the description of quantum states and their properties. With
reference to group $S_{4}$ it should be noted that among its 30 subgroups
there are 4 subgroups that are isomorphic to group $S_{3}$ (that is
permutation group of 3 elements).Exactly this subgroup can serve as
demonstrative and instructive example of the small (and entangled as we see
further) quantum world. Let us consider now the concrete realization of this
subgroup which leaves as invariant the fourth element (state) of the group
and construct the model of SQW based on this realization. In the standard
basis of two-qubit states the relevant algebra of observables acting in this
SQW consists of six generators, from which three (aside from unit operator)
are Hermitian and the rest two are unitary. Let us write down them in
explicit form. The three Hermitian operators are:
\begin{eqnarray}
{\hat{H}_{1}} &=&%
\begin{pmatrix}
0 & 1 & 0 & 0 \\
1 & 0 & 0 & 0 \\
0 & 0 & 1 & 0 \\
0 & 0 & 0 & 1%
\end{pmatrix}%
,{\hat{H}_{2}}=%
\begin{pmatrix}
0 & 0 & 1 & 0 \\
0 & 1 & 0 & 0 \\
1 & 0 & 0 & 0 \\
0 & 0 & 0 & 1%
\end{pmatrix}%
,  \notag  \label{2} \\
{\hat{H}_{3}} &=&%
\begin{pmatrix}
1 & 0 & 0 & 0 \\
0 & 0 & 1 & 0 \\
0 & 1 & 0 & 0 \\
0 & 0 & 0 & 1%
\end{pmatrix}%
\end{eqnarray}%
and two unitary ones are
\begin{equation}
\hat{A}=%
\begin{pmatrix}
0 & 1 & 0 & 0 \\
0 & 0 & 1 & 0 \\
1 & 0 & 0 & 0 \\
0 & 0 & 0 & 1%
\end{pmatrix}%
\text{ , }{\hat{B}}=%
\begin{pmatrix}
0 & 0 & 1 & 0 \\
1 & 0 & 0 & 0 \\
0 & 1 & 0 & 0 \\
0 & 0 & 0 & 1%
\end{pmatrix}%
.  \label{3}
\end{equation}

Note that unitary operators $\hat{A}$ and ${\hat{B}}$ are conjugate and
reciprocal to each other, that is $\hat{A}={\hat{B}^{+}}$ and $\hat{A}\hat{B}%
={\hat{B}}\hat{A}=\hat{1}$. Let us point out the complete system of
relations for hermitian generators of the algebra: ${\hat{H}_{i}}^{2}=\hat{1}
$ ($i=1,2,3$), ${\hat{H}_{1}}{\hat{H}_{2}}={\hat{H}_{2}}{\hat{H}_{3}}={\hat{H%
}_{3}}{\hat{H}_{1}}=\hat{A}$ and $\hat{H_{1}}{\hat{H}_{3}}={\hat{H}_{2}}{%
\hat{H}_{1}}={\hat{H}_{3}}\hat{H_{2}}={\hat{B}}$. In addition let us give
also the algebraic relations connecting operators ${\hat{H}_{i}}$ with
unitary operators $\hat{A}$ \ which have the next form: ${\hat{H}_{1}}\hat{A}%
={\hat{H}_{2}},$ ${\hat{H}_{2}}\hat{A}={\hat{H}_{3}},$ $\hat{H_{3}}\hat{A}={%
\hat{H}_{1}}$ and $\hat{A}{\hat{H}_{1}}=\hat{H_{3}},\hat{A}{\hat{H}_{2}}={%
\hat{H}_{1}},\hat{A}{\hat{H}_{3}}={\hat{H}_{2}}$ .The similar equations
including the matrix ${\hat{B}}$ can be obtained from these relations by
conjugation. There are also two relations, connecting operators $\widehat{A}$
and $\widehat{B\text{ }}$ : $\hat{A}^{2}={\hat{B}}$,and ${\hat{B}}^{2}=\hat{A%
}$. The density matrix of any state belonging to this SQW may be represented
as $\hat{\rho}=\frac{k}{2}\hat{1}+l{\hat{H}_{1}}+m{\hat{H}_{2}}+n\hat{H_{3}}%
+p\left( \hat{A}+{\hat{B}}\right) $, where $k,l,m,n,p$ are real numbers
satisfying the normalization condition: $k+l+m+n+p=\frac{1}{2}$. However, it
is easy to see that generators of the algebra ${\hat{H}_{i}}$, $\hat{A}$ and
${\hat{B}}$ are connected by the additional relation : $\hat{A}+{\hat{B}}=%
\hat{C}-1$, where $\hat{C}\equiv {\hat{H}_{1}}+{\hat{H}_{2}}+{\hat{H}_{3}}$.
Thus the general representation for the density matrix in given SQW may be
written finally as :
\begin{equation}
\hat{\rho}=\frac{a}{2}+b{\hat{H}_{1}}+c{\hat{H}_{2}}+d{\hat{H}_{3}}
\label{4}
\end{equation}%
with normalization condition: $a+b+c+d=\frac{1}{2}$.

It is worth noting that operator $\hat{C}$ commutes with all
generators of the algebra and hence is the Casimir operator of the
group $S_{3}$. Let us turn now to the question of clarifying the
structure of pure states in this SQW. Starting from the
representation \eqref{4} and using the defining condition for pure
states: ${\hat{\rho}}^{2}={\hat{\rho}}$, one, after elementary
calculations, can obtain the following restrictions on the
coefficients of the decomposition \eqref{4}, namely$:$
\begin{eqnarray}
&&\frac{a^{2}}{2}=\frac{a^{2}}{4}+b^{2}+c^{2}+c^{2}-(bc+bd+cd);  \notag
\label{5} \\
&&b=ab+(bc+bd+cd);  \notag \\
&&c=ac+(bc+bd+cd);  \notag \\
&&d=ad+(bc+bd+cd).
\end{eqnarray}

It is clear that the only possibility to satisfy all relations \eqref{5} is
to put the coefficient $a$ equal to unit and impose on the coefficients $%
b,c,d$ the next restriction: $b^{2}+c^{2}+d^{2}=\frac{1}{4}$ (together with
normalization condition $b+c+d=-\frac{1}{2}$). Thus, in the parameter space
of coefficients $b,c,d$ the set of pure states is the intersection of the
sphere centered in the origin, whose radius is equal to $\frac{1}{2}$, and
the plane satisfying the equation $b+c+d=-\frac{1}{2}$. Evidently it is a
circle. Further, it is well-known that in Hilbert space pure states form the
boundary of a convex set of all quantum states \cite{3}. Hence the result
obtained means that all mixed states of the system in this SQW must be
settled within the circle specified above. Thus the structure of quantum
states in the SQW under study is quite simple and obvious. In addition one
can point out \ the parametrization of pure states in this SQW by writing
the coefficients of \eqref{4} with $a=1$ in the next convenient form:
\begin{eqnarray}
b &=&-\frac{t\left( 1+t\right) }{2\left( 1+t+t^{2}\right) };c=-\frac{\left(
1+t\right) }{2\left( 1+t+t^{2}\right) };  \notag  \label{6} \\
d &=&\frac{t}{2\left( 1+t+t^{2}\right) }
\end{eqnarray}%
(with the only real parameter $t).$ It is easy to verify directly that two
relations $b+c+d=-\frac{1}{2}$ and $b^{2}+c^{2}+d^{2}=\frac{1}{4}$
characterizing pure states in this SQW are fulfilled automatically. Using
the parametrization \eqref{6} one can write down the normalized vector $%
\left\vert \Psi \right\rangle $ corresponding to the density matrix of a
pure state, that is, if ${\hat{\rho}}=\left\vert \Psi \right\rangle
\left\langle \Psi \right\vert $, then the appropriate vector $\left\vert
\Psi \right\rangle $ can be represented as: $\left\vert \Psi \left( t\right)
\right\rangle =\frac{1}{\sqrt{2\left( 1+t+t^{2}\right) }}%
\begin{pmatrix}
1+t \\
t \\
1 \\
0%
\end{pmatrix}%
.$

Since all information contained in quantum state of the system can be
extracted only by making appropriate measurements , it is useful to give a
simple experimental criterion of the state purity. To this end we assume
that the unknown quantum state has the above-mentioned form:
\begin{multline}
{\hat{\rho}}=\frac{1}{2}+b{\hat{H}_{1}}+c{\hat{H}_{2}}+d\hat{H_{3}}=
\label{7} \\
=%
\begin{pmatrix}
\frac{1}{2}+d & b & c & 0 \\
b & \frac{1}{2}+c & d & 0 \\
c & d & \frac{1}{2}+b & 0 \\
0 & 0 & 0 & 0%
\end{pmatrix}%
.
\end{multline}

First of all let us find the eigenvalues of the density matrix \eqref{7}.
One can verify easily that ${\hat{\rho}}$ has two zero eigenvalues: the
first with eigenvector $\left\vert 0\right\rangle _{1}=%
\begin{pmatrix}
0 \\
0 \\
0 \\
1%
\end{pmatrix}%
$ and the second with eigenvector $\left\vert 0\right\rangle _{2}=\frac{1}{%
\sqrt{3}}%
\begin{pmatrix}
1 \\
1 \\
1 \\
0%
\end{pmatrix}%
$. Besides them there are two nonzero eigenvalues $\mu _{1}$and $\mu _{2}$
that satisfy to the quadratic equation: $\mu ^{2}-\mu +3\left(
bc+bd+cd\right) =0$ with the solutions:
\begin{equation}
\mu _{1,2}=\frac{1\pm \sqrt[2]{1-12\left( bc+bd+cd\right) }}{2}  \label{8}
\end{equation}%
The expression \eqref{8} implies that coefficients $b,c,d$ aside from
normalization condition must satisfy to the inequality $0\leq \left(
bc+bd+cd\right) \leq \frac{1}{12}.$

Let us define now as usual the mean value of an arbitrary observable $\hat{A}
$ as $\left\langle \hat{A}\right\rangle =\mathrm{Tr}\left( \hat{ \rho } \hat{%
A}\right)$. Using the representation \eqref{7} one can write down the mean
values of the following three selected observables:

1) $\left\langle H_{1}\right\rangle -1\equiv A_{1}=c+d+4b,$

2) $\left\langle H_{2}\right\rangle -1\equiv A_{2}=b+d+4c,$ and

3) $\left\langle H_{3}\right\rangle -1\equiv A_{3}=b+c+4d.$

Taking into account the condition $b+c+d=-\frac{1}{2}$ one can find that $\
\ \ \left\langle A_{1}+A_{2}+A_{3}\right\rangle =-3$, that is for all states
of SQW (with $a=1$) the mean value of the Casimir operator $\left\langle
C\right\rangle \equiv \left\langle H_{1}+H_{2}+H_{3}\right\rangle =0$. On
the other hand if one considers another useful quantity, namely $R\equiv
A_{1}^{2}+A_{2}^{2}+A_{3}^{2}=\left( c+d+4b\right) ^{2}+\left( b+d+4c\right)
^{2}+\left( b+c+4d\right) ^{2}$, then she(he) obtains that $R=\left( -\frac{1%
}{2}+3b\right) ^{2}+\left( -\frac{1}{2}+3c\right) ^{2}+\left( -\frac{1}{2}%
+3d\right) ^{2}=9\left( \frac{1}{4}+b^{2}+c^{2}+d^{2}\right) =\frac{9}{2}$.
Thus, for all pure states in SQW the following criterion of purity should be
true:
\begin{equation}
\left( \left\langle H_{1}\right\rangle -1\right) ^{2}+\left( \left\langle
H_{2}\right\rangle -1\right) ^{2}+\left( \left\langle H_{3}\right\rangle
-1\right) ^{2}=\frac{9}{2}  \label{9}
\end{equation}%
Let us go to the next question which we are interested in: how the
explicit expression for the entanglement of the states (both pure
and mixed) in this SQW looks? For the sake of simplicity as before
we are limited ourselves to studying the states for which the
coefficient $a$ in \eqref{4} is equal to unit and the appropriate
density matrix can be represented as
\begin{equation}
{\hat{\rho}}=\frac{1}{2}+b{\hat{H}_{1}}+c{\hat{H}_{2}}+d{\hat{H}_{2}}
\label{10}
\end{equation}%
with a normalization condition: $b+c+d=-\frac{1}{2}$. As was explained
above, among these states there are certain pure states (for which $%
b^{2}+c^{2}+d^{2}=\frac{1}{4}$) while all the rest are mixed. Let us
determine the entanglement of the state \eqref{10}. To this end in the case
of two-qubit arbitrary mixed state the well-known recipe was proposed by
Wootters in \cite{4}. This recipe reads as follows. First of all one must
construct the auxiliary matrix $\tilde{\rho}=\left( \sigma _{y}\otimes
\sigma _{y}\right) \rho ^{\ast }\left( \sigma _{y}\otimes \sigma _{y}\right)
$, where $\sigma _{y}$ $=%
\begin{pmatrix}
0 & -i \\
i & 0%
\end{pmatrix}%
$ is the Pauli matrix and ${\hat{\rho}^{\ast }}$ is the matrix
conjugated to given matrix ${\hat{\rho}} $ \eqref{10}. At the next
step one needs
to introduce another auxiliary matrix $\hat{\Omega}$ $=\hat{\rho}{\hat{\rho}}%
^{\ast }$ (that is non-Hermitian and positive) and after that to find its
four eigenvalues $\Omega _{i}$ ($i=1,2,3,4$). If one arranges them in
decreasing order $\Omega _{1}\geq \Omega _{2}\geq \Omega _{3}\geq \Omega
_{4}\geq 0$, then according to the paper \cite{4} the entanglement of
formation $E(\widehat{\rho })$ for the state ${\hat{\rho}}$ $\ $ may be
calculated as follows:
\begin{multline}
E\left( {\hat{\rho}}\right) =-\frac{\left( 1+\sqrt{1-C^{2}}\right) }{2}\log
_{2}\frac{1+\sqrt{1-C^{2}}}{2}-  \label{11} \\
-\frac{\left( 1-\sqrt{1-C^{2}}\right) }{2}\log _{2}\frac{\left( 1-\sqrt{%
1-C^{2}}\right) }{2},
\end{multline}%
where the concurrence
\begin{equation*}
C=C\left( {\hat{\rho}}\right) =\max \left\{ 0,\sqrt{\Omega _{1}}-\sqrt{%
\Omega _{2}}-\sqrt{\Omega _{3}}-\sqrt{\Omega _{4}}\right\} .
\end{equation*}%
Note that since $E\left( {\hat{\rho}}\right) $ is a monotonic and increasing
function of concurrence $C\left( \widehat{\rho }\right) $, we can restrict
ourselves to determination of the value of $C\left( \widehat{\rho }\right) $
that evidently ranges from zero to unit and defines the degree of
entanglement as well as $E\left( {\hat{\rho}}\right) $. Omitting elementary
calculations we give the final expressions for the auxiliary matrix $\hat{%
\Omega}$ and its four eigenvalues, namely:
\begin{equation}
\hat{\Omega}=%
\begin{pmatrix}
0 & b\left( \frac{1}{2}+b\right) +cd & bd+c\left( \frac{1}{2}+c\right) & -2bc
\\
0 & \left( \frac{1}{2}+b\right) \left( \frac{1}{2}+c\right) +d^{2} & 2\left(
\frac{1}{2}+c\right) d & 0 \\
0 & 2\left( \frac{1}{2}+b\right) d &  & 0 \\
0 & 0 & 0 & 0%
\end{pmatrix}
\label{12}
\end{equation}%
The eigenvalues of matrix $\hat{\Omega}$ (in decreasing order) are equal to:
$\Omega _{1}=\left[ d+\sqrt{\left( \frac{1}{2}+b\right) \left( \frac{1}{2}%
+c\right) }\right] ^{2},\Omega _{2}=\left[ d-\sqrt{\left( \frac{1}{2}%
+b\right) \left( \frac{1}{2}+c\right) }\right] ^{2},\Omega _{3}=\Omega
_{4}=0.$

Using the above-mentioned recipe for determination of entanglement one can
find the required result:
\begin{equation}
C\left( {\hat{\rho}}\right) =2\sqrt{\left( \frac{1}{2}+b\right) \left( \frac{%
1}{2}+c\right) }.  \label{13}
\end{equation}%
It is worth noting that if the state \eqref{6} is pure, that is ${\hat{\rho}}%
=\left\vert \Psi \right\rangle \left\langle \Psi \right\vert $, where $%
\left\vert \Psi \right\rangle =\frac{1}{2\sqrt{1+t+t^{2}}}%
\begin{pmatrix}
1+t \\
t \\
1 \\
0%
\end{pmatrix}%
$, then the expression \eqref{13} takes the form: $C\left\{ \Psi \right\} =%
\frac{\left\vert t\right\vert }{1+t+t^{2}}$ which coincides with the
standard definition of concurrence of a pure state$\left\vert \Psi
\right\rangle $. We see that unlike of general case the entanglement
of states belonging to the SQW considered may be explicitly
expressed in terms of the coefficients of the given density matrix
only.Now we turn to the study another interesting problem, namely,
to clarify how much entanglement may be extracted from the given
state (pure or mixed) belonging to the SQW by means of various
measurements carried out on a system being in the state \eqref{10}.
Clearly, having in hands the simple expression for the amount of
entanglement contained in any quantum state of SQW \eqref{13}, this
problem may be easily solved. We consider here only the cases when
the measured observables are the basic observables that is
generators of the algebra $H_{1},H_{2},H_{3}$

Note that in view of relation ${\hat{H}_{i}}^{2}=1\left( \text{for all }%
i=1,2,3\right) $ one can write a simple equation connecting two density
matrices,namely,density matrix ${\hat{\rho}_{0}}$ before the measurement and
the density matrix ${\hat{\rho}_{\infty }}$ after the measurement of the
observable ${\hat{H}_{i}}$. This equation reads as:
\begin{equation}
{\hat{\rho}_{\infty }}=\frac{{\hat{\rho}_{0}}+{\hat{H}_{i}}{\hat{\rho}_{0}}{%
\hat{H}_{i}}}{2}.  \label{14}
\end{equation}%
Using Eq.\eqref{14} and the algebra of operators described above one can
consider separately three cases: the case I when the observable ${\hat{H}_{1}%
}$ \ is measured and the initial state is ${\hat{\rho}_{0}}=\frac{1}{2}+b{%
\hat{H}_{1}}+c{\hat{H}_{2}}+d{\hat{H}_{3}}$, the case II when the observable
${\hat{H}_{2}}$ is measured and the case III when observable ${\hat{H}_{3}}$
is measured. In the case I after the measurement the state of the system is:
${\hat{\rho}_{\infty }}^{I}=\frac{1}{2}+b{\hat{H}_{1}}+\frac{(c+d)}{2}\left(
{\hat{H}_{2}}+{\hat{H}_{3}}\right) $. Similarly in the case II the final
state is ${\hat{\rho}_{\infty }}^{II}=\frac{1}{2}+\frac{(b+d)}{2}\left( \hat{%
H_{1}}+{\hat{H}_{3}}\right) +c{\hat{H}_{2}}$ and in the case III the final
stateof the system after measurement is ${\hat{\rho}_{\infty }}^{III}=\frac{1%
}{2}+\frac{\left( b+c\right) }{2}\left( {\hat{H}_{1}}+{\hat{H}_{2}}\right) +d%
{\hat{H}_{3}}$. Now we are interested in the maximum amount of entanglement
that can be extracted from initial state by these different measurements.
Let us calculate this maximum. Without loss of generality we may assume that
the initial state of the system is pure and can be represented by
parametrization \eqref{6}. Then the gain of entanglement caused by
measurement of ${\hat{H}_{1}}$ may be written as:
\begin{multline}
\Delta C_{I}=C\left\{ {\hat{\rho}_{\infty }}^{I}\right\} -C\left\{ {\hat{\rho%
}_{0}}\right\} =  \label{15} \\
=2\sqrt{\left( b+\frac{1}{2}\right) \left( \frac{c+d+1}{2}\right) }-2\sqrt{%
\left( b+\frac{1}{2}\right) \left( c+\frac{1}{2}\right) }= \\
=\frac{1}{1+t+t^{2}}\left[ \sqrt{\frac{1+2t+2t^{2}}{2}}-\left\vert
t\right\vert \right] .
\end{multline}%
Note that in the derivation of relation \eqref{15} we used the
parametrization \eqref{6} for coefficients $b,c,d$ of the initial pure
state. It is easy to see that maximum \eqref{15} is equal to $\frac{1}{\sqrt{%
2}}$ and is reached when $t=0$. The required initial state in this case is $%
\left\vert \Psi _{0}\right\rangle _{I}=\frac{1}{\sqrt{2}}%
\begin{pmatrix}
1 \\
0 \\
1 \\
0%
\end{pmatrix}%
$. In the same way one can find that in the case II the gain of entanglement
can be represented as:
\begin{equation}
\Delta C_{II}=\frac{\left\vert t\right\vert }{1+t+t^{2}}\left[ \sqrt{\frac{%
2+2t+t^{2}}{2}}-1\right] .  \label{16}
\end{equation}%
The maximum of \eqref{16} is reached when $t=\pm \infty $ and is equal to $%
\frac{1}{\sqrt{2}}$ as well. The required initial state in this case is $%
\left\vert \Psi _{0}\right\rangle _{II}=\frac{1}{\sqrt{2}}%
\begin{pmatrix}
1 \\
\pm 1 \\
0 \\
0%
\end{pmatrix}%
$. However, in the case III, when the observable ${\hat{H}_{3}}$ is
measured, the result is somewhat distinct, namely, he gain of entanglement
caused by this measurement can be written as:
\begin{equation}
\Delta C_{III}=C\left\{ {\hat{\rho}_{\infty }}^{III}\right\} -C\left\{ {\hat{%
\rho}_{0}}\right\} =\frac{1+t^{2}-2\left\vert t\right\vert }{2\left(
1+t+t^{2}\right) }.  \label{17}
\end{equation}%
It is easy to see that the maximum of \eqref{17} is reached when $t=0$ and
is equal to $\frac{1}{2}$. It is curious that, although optimal initial
states for cases I and III coincide, nevertheless, the extracted amount of
entanglement is larger in the first case. Now let us consider another
important and interesting feature of certain states belonging to the SQWs
that makes them potentially very helpful in various quantum informational
applications. We have in mind the existence of irreducible entangled states
both in SQW under examination and in many others SQWs as well. Really let us
consider the selected mixed state $IE$ (that is irreducible and entangled)
with density matrix ${\hat{\rho}}_{IE}=\frac{1}{2}-\frac{\hat{C}}{6}\equiv
\frac{1}{2}-\frac{\hat{H_{1}}+{\hat{H}_{2}}\text{+}{\hat{H}_{3}}}{6}$. It is
easy to see that this state belongs to the SQW because two necessary
conditions: $b+c+d=-\frac{1}{2}$ and $bc+bd+cd=\frac{1}{12}\leq \frac{1}{12}$%
are satisfied. On the other hand, it is clear that this state is irreducible
(that is cannot be changed in time by any dynamical way or by means of
measurements) because the Casimir operator $\hat{C}={\hat{H}_{1}}+{\hat{H}%
_{2}}+{\hat{H}_{3}}$ commutes with all generators of the algebra.

In addition note that this selected state is entangled with concurrence $%
C_{IE}=\frac{2}{3}$. Thus we come to the conclusion that similar states ( in
the case when realization of the SQW would be possible) may be used as
long-lived keepers of entanglement stored in the system. The mentioned
feature of irreducible entangled states makes them indispensable for various
quantum informational applications. It should also be noted that although
the IE state is dynamically stable it can be achieved easily from other
states by means of appropriate measurements. For example if one takes the
initial state of the system in the form: ${\hat{\rho}_{0}}=\frac{1}{2}-\frac{%
{\hat{H}_{1}}}{6}+c{\hat{H}_{2}}+d{\hat{H}_{3}}$ and then performs the
measurement of the observable $\hat{H_{1}}$ in this state then (if the
conditions: 1) $c+d=-\frac{1}{3}$ and 2) $cd\leq \frac{5}{36}$ hold true)
she(he) gets exactly the required IE state after the measurement.

Let us sum up the main results obtained in the present paper. We introduce
the notion of small quantum world (SQW) which allows one to study special
classes of composite quantum systems whose pure states and nonlocal
properties turn out to be simpler by far comparing with large quantum worlds
containing them as a small part. We are demonstrating that some features of
selected states belonging to these SQWs would be very helpful for performing
various quantum information tasks.

\end{document}